\documentclass[preprint,review,12pt]{elsarticle}
\journal{International Journal of Engineering Science}
\usepackage{amssymb}
\usepackage{float}
\usepackage{amsmath}
\usepackage{graphicx}
\usepackage{bm}
\usepackage{color}
\DeclareMathAlphabet{\mathpzc}{OT1}{pzc}{m}{it}
\newcommand{\uu}{\bm u}                        % displacement vector
\newcommand{\vel}{\bm v}                       % velocity vector
\newcommand{\stress}{\bm\sigma}                % local stress tensor
\newcommand{\Stress}{\bm\Sigma}                % non-local stress tensor
\newcommand{\strain}{\bm\varepsilon}           % strain
\newcommand{\f}{\bm f}                         % body force
\newcommand{\curr}{\bm J}                      % carbon current
\newcommand{\power}{\mathcal P}                % power
\newcommand{\powerden}{\mathpzc p}             % power density
\newcommand{\ie}{e}                            % internal energy
\newcommand{\ent}{\eta}                        % entropy
\newcommand{\fe}{\psi}                         % free energy density
\newcommand{\Fe}{\Psi}                         % free energy
\newcommand{\q}{\bm q}                         % heat flux
\newcommand{\bracs}[1]{\left[ {#1} \right]}    % square brackets [.]
\newcommand{\vect}[1]{\boldsymbol{#1}}         % vector

\let\oldmarginpar\marginpar
\renewcommand\marginpar[1]{\-\oldmarginpar[\raggedright\scriptsize #1]{\raggedright\scriptsize #1}}

\begin{document}
\begin{frontmatter}
\title{A thermodynamical model for concurrent diffusive and displacive phase
transitions}
\author[unibo1]{Mirko Maraldi \corref{cor1}}
\cortext[cor1]{Corresponding author.}
\ead{mirko.maraldi@unibo.it}
\author[unibo2]{Luisa Molari}
\author[unibo3]{Diego Grandi}
\address[unibo1]{DIEM, Universit\`a di Bologna, v.le Risorgimento 2, 40136 Bologna, Italy}
\address[unibo2]{DISTART, Universit\`a di Bologna, v.le Risorgimento 2, 40136 Bologna, Italy}
\address[unibo3]{DM, Universit\`a di Bologna, Piazza di Porta S. Donato 5, 40127 Bologna, Italy}
\begin{abstract}
A thermodynamically consistent framework able to model either diffusive and
displacive phase transitions is proposed. The first law of thermodynamics, the
balance of linear momentum equation and the Cahn-Hilliard equation for solute mass
conservation are the governing equations of the model, which is complemented by
a suitable choice of the Helmholtz free energy and consistent boundary and initial
conditions. Some simple test cases are presented that demonstrate the potential
of the model to describe both diffusive and displacive phase transitions as well
as the related changes in temperature.
\end{abstract}
\end{frontmatter}
%%%%%%%%%%%%%%%%%%%%%%%%%%%%%%%%%%%%%%%%%%%%%%%%%%%%%%%%%%%%%%%%%%%%%%%%%%%%%%%%

\section{Introduction} \label{intro}
\begin{figure}
 	\footnotesize
 	\centering
  \begin{tabular}{ccc}
    \includegraphics[height=3.8cm]{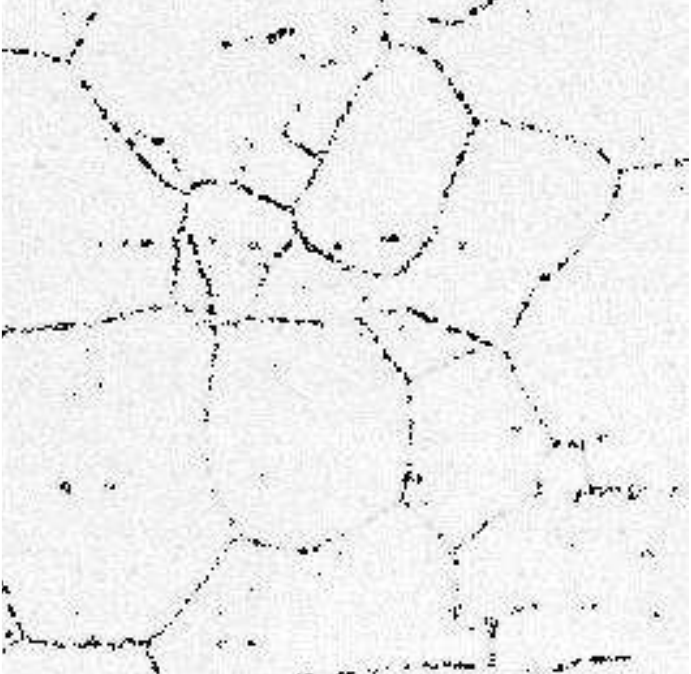} &
    \includegraphics[height=3.8cm]{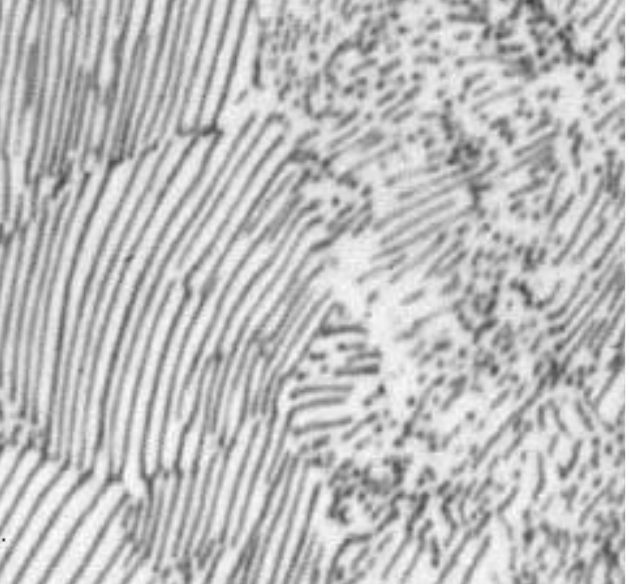}  &
    \includegraphics[height=3.8cm]{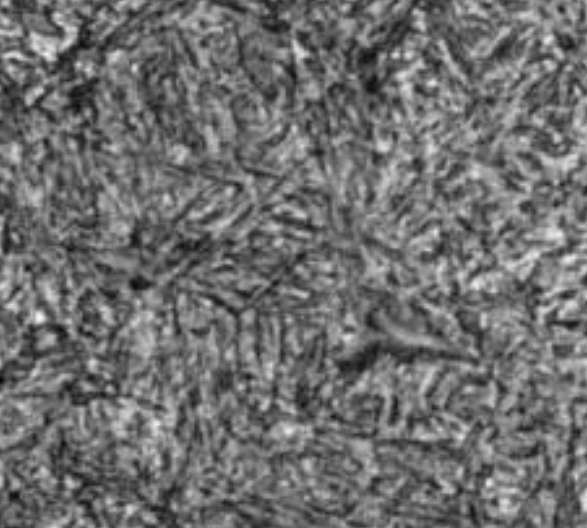} \\
 		(a)  & (b) & (c)
 	\end{tabular}
 	\caption{Microscope image of: (a) austenite, (b) perlite and (c)
     martensite \cite{asm}.}
  \label{fig:Micros}
\end{figure}

Solid-to-solid phase transitions can be of different nature: diffusive and
displacive. Diffusive transformations are second order phase transitions, and
are able to describe, among others, processes involving diffusion of atoms. Typical
examples include spinodal decomposition, vapour-phase deposition, crystal growth
during solidification and grain growth in single-phase and two-phase systems.
Spinodal decomposition can occur, for instance, in an eutectoid steel: a
diffusive transition from a high temperature stable phase, austenite
(Fig.\ref{fig:Micros}a), to perlite (Fig.\ref{fig:Micros}b) takes place at low
cooling rate; as the temperature decreases slowly, carbon (the solute) can
migrate outside iron cell and originate cementite layers.

Displacive transformations are first order phase transitions involving a lattice
distortion. Typical examples include the martensitic transformation of shape
memory alloys or martensitic transformation of quenched steels: here the
displacive transformation from austenite to martensite (Fig.\ref{fig:Micros}c)
occurs at high cooling rate; the temperature decreases fast, and there is
no time for diffusive phenomena to take place so that the new phase originates
with a lattice deformation.

Diffusive and displacive phase transitions are concurrently involved in many and
very important industrial processes, such as the heat treatments of steel. In
particular, in a steel subjected to a heat treatment there may be different zones
in which, due to a different cooling rate, the transformations are of different
kind, leading to different phases.

A number of approaches may be adopted to model both the aforementioned phase
transitions; for instance, use of phenomenological laws can be made: Brokate
\cite{brokate}, following the approach proposed by Hoemberg \cite{hoemberg},
presents a model constituted by the Koistinen-Marburger rule for the prediction
of the martensitic phase fraction, the Johnson-Mehl formula for the estimate of
the perlitic phase fraction, together with the Scheil's additivity rule for
non-isothermal conditions. However, in Author's opinion, such an approach does not
make clear the physics behind the phenomena involved. A different way to tackle
the problem may consist in setting a framework in the context of nonlinear
thermoelasticity including the Ginzburg-Landau theory of phase transitions
\cite{landau}, \cite{toledano}. Following the phase-field approach, a microstructure
is identified by a variable named order parameter. The phase transitions rely on
phase evolution equations; constitutive laws are expressed as the derivatives of
the Helmholtz free energy density, which is given as a function of the order
parameters.

In particular, a Ginzburg-Landau theory was proposed by Falk \cite{falk} for the
martensitic phase transition in shape memory alloys using an order parameter
dependent on the strain tensor. The same idea has been developed by Barsch et.
al. \cite{barsch} to describe the microstructure of inhomogeneous materials and
their proposed model has been extended to simulate martensitic structures in two-
and three-dimensional domains. In the work by Ahluwalia et. al. \cite{ahluwalia}
this theory is generalised to describe a two-dimensional square to rectangle
martensitic transition in a polycrystal with more than one lattice orientation.
Other models which describe the martensitic transformation in a Ginzburg-Landau
framework using a different order parameter not dependent on the strain tensor
are proposed by Levitas et. al.\cite{levitas}, Wang et. al.\cite{WK}, Artemev et.
al.\cite{Artemev2}, Berti et. al.\cite{berti}.

Ginzburg-Landau frameworks have also been applied to study diffusive phase
transitions. In this context, Cahn \cite{Cahn00} presented a first model in the
simplest case of diffusive phase transition without accounting for thermal
effects. An extension to Cahn's model is presented - among others - by Alt et.
al. \cite{alt}; in their paper the coupled phenomena of mass diffusion and heat
conduction in a binary system subjected to thermal activation have been modelled.
Other models accounting for the mechanical aspects related to the diffusive phase
transitions are the ones by Onuki et. al. \cite{onukifu} and Fried et. al.\cite{fried}.

First attempts to study diffusive and displacive phase transitions together were
made by Rao et. al. \cite{rao} and by Levitas \cite{levitas2}; the same idea has
been developed by Bouville et. al. \cite{bouville07}, who proposed a Ginzburg-Landau
framework including the balance of linear momentum equation and the Cahn-Hilliard
equation and analysed the effects of a volume change consequent to a diffusive
or a displacive phase transformation.

In this paper, an attempt to describe both displacive and diffusive phase
transitions in a thermodynamically consistent framework is made; the coupling
between thermal, chemical and mechanical effects is accounted for, and a way to
overcome the difficulties arising from the treatment of the gradient terms is
proposed. Three are the governing equations, presented as balance equations for
observable variables: the balance of linear momentum, the Cahn-Hilliard equation
as a solute mass balance \cite{cahn} and the heat equation (balance of internal
energy). The model is completed with a suitable description of the free energy.
The analysis is reduced to a two dimensional setting, which is simpler than a
three dimensional one, but still meaningful.
% Some simple test cases are solved to complement the theoretical formulation of
% the model.
%

The outline of the paper is as follows: after explaining the choice of the order
parameters for each transformation in Sect.\ref{OPs}, in Sect.\ref{balance}
balance equations are provided for linear momentum, solute mass and internal
energy; the first two will act as the evolution equations needed for a
Ginzburg-Landau framework. In Sect.\ref{second-law} the restrictions provided by
the second law of thermodynamics are highlighted and its exploitation in the
form of the Clausius-Duhem inequality is provided. In Sect.\ref{free energy} the
model is completed with the description of the Helmholtz free energy density, which
will allow an explicit form for the constitutive laws to be derived in
Sect.\ref{explicit}. The resulting model is a diffuse interface one, and this
implies the presence of the gradient of the order parameters, which will put some
issues regarding the exploitation of the second principle of thermodynamics
(Sect.\ref{second-law}) and the necessary high order boundary conditions, as
highlighted in Sect.\ref{BCs}.
% In Sect.\ref{simulations} the main features of the model are illustrated by
% means of three simple test cases. In the first two tests the diffusive and
% the displacive phase transitions are studied separately and the peculiar
% characteristics of each transition are highlighted; the third test is instead
% devoted to the analysis of the interplay between the different phases the
% model accounts for.
The paper ends drawing some conclusions.
%%%%%%%%%%%%%%%%%%%%%%%%%%%%%%%%%%%%%%%%%%%%%%%%%%%%%%%%%%%%%%%%%%%%%%%%%%%%%%%%

\section{Model of the problem}
\label{OPs}
The model is set at a microscopic scale of observation which may be, for instance,
a single-grain scale and relies on a number of order parameters to describe the
different kinds of transformations involved. In order to account for the presence
of three phases (austenite, perlite and martensite), two order parameters
playing as the unknowns of the problem may be introduced.

%-------------------------------------------------------------------------------
\subsection{Diffusive order parameter}
\label{sec:diffusive_op}
To account for diffusive phenomenon, an order parameter called $c$ is set:
\begin{equation}
  \label{eqn:op_diffusive}
  c = c ({\bm x}, t), \quad   {\rm for} \
	  ({\bm x}, t) \in \Omega \  \times \ [0, \infty),
\end{equation}
being $\Omega$ a two-dimensional domain.

This parameter describes the evolution from one phase to another one with a
different structure. According to experimental observations and the use of the
Cahn-Hilliard equation \cite{cahn} (see Sect.\ref{balance}), the order parameter
may represent the solute mass fraction. Hence, Pearlite, and in general phases
obtained by diffusive transformations, will be the one for which this order
parameter is non-zero, i.e. the carbon has migrate to generate areas in which
there is a great quantity $(c > 0)$ in spite of other areas, in which there is a
lack of carbon $(c < 0)$. Moreover, the higher value $c$ will assume in a point
of the domain, the more the transformation will be at a late stage there. Given
this interpretation, the Cahn-Hilliard equation may be seen as a balance of mass
equation.

%------------------------------------------------------------------------------
\subsection{Displacive order parameter}
\label{sec:displacive_op}
Regarding martensitic phase formation, a shear strain needs to be applied to the
lattice in order to obtain the phase change \cite{lieberman}, \cite{bhadeshia}.
For this reason, in considering displacive phenomena, the common choice for the
order parameter, called $e_{2}$, in a two-dimensional setting \cite{ahluwalia}
\cite{shenoy} \cite{albers} is
\begin{equation}
  e_{2} = \frac{\varepsilon_{11}-\varepsilon_{22}}{\sqrt{2}},
  \label{eqn:diffusive_op2}
\end {equation}
where
\begin{equation}
  {\strain} = \frac{1}{2} \left( \nabla{\bm u}+ \nabla {\bm u}^T \right)
\end{equation}
is the linearised strain tensor and
\begin{equation}
  \label{eqn:op_displacive}
  \bm u = \bm u ({\bm x}, t), \quad   {\rm for} \
	  ({\bm x}, t) \in \Omega \  \times \ [0, \infty)
\end{equation}
is the displacement field.

%%%%%%%%%%%%%%%%%%%%%%%%%%%%%%%%%%%%%%%%%%%%%%%%%%%%%%%%%%%%%%%%%%%%%%%%%%%%%%%%

\section{Balance equations}
\label{balance}
We start the description of the model by stating the balance equations and then
by examining the restrictions imposed by the second law of thermodynamics. We
will last enter appropriate constitutive hypothesis by choosing a constitutive
law for the free energy.

The model consists of three balance equations, reported here in their pointwise
formulation:
\begin{itemize}
  \item balance of linear momentum:
  \begin{equation}\label{eqn:bal-momentum}
    \rho\ddot\uu - \nabla\cdot\stress = \rho \f,
  \end{equation}
    where $\stress$ is the stress tensor and $\f$ is the external body force;
  \item balance of solute mass:
  \begin{equation}\label{eqn:bal-carbon}
     \dot c = - \nabla \cdot \curr,
  \end{equation}
  where $\curr = c \bm V$ is the flux vector related to the solute mass
  fraction ($\bm V$ is its the velocity);
  \item balance of energy:
  \begin{equation}\label{eqn:bal-e}
     \dot\ie=\powerden^i_m + \powerden^i_c+h;
  \end{equation}
  here, $\ie$ is the internal energy density, $\powerden^i_m$  is the mechanical
  internal power (density), $\powerden^i_c$  is the internal power associated
  to the  solute mass balance equation and $h$ is the internal thermal power.
  This can be expressed by the thermal power balance
  \begin{equation}
     h = - \nabla \cdot \q + r,
  \end{equation}
  $\q$ being the heat flux and $r$ the external heat supply.
\end{itemize}
The balance of linear momentum and of solute mass give rise to two power balance
equations.

To account for non-local effects and spatial variations of the order parameters
(interfaces), we demand a non-local constitutive relation for stress and chemical
potential. Some cares are needed due to the presence of non-localities in the
constitutive equations, as this will lead to non-conventional expressions for the
internal powers associated to momentum and solute mass balance.

For the time being it suffices to consider this relation in the quite generic
form
\begin{equation}
  \stress = \stress_{loc}(\Gamma_{loc},\dot\Gamma_{loc}) - \nabla\cdot\Stress(\Gamma),
\end{equation}
where $\Gamma_{loc} = (\strain, c, T)$ are the local state variables and
$\Gamma = (\Gamma_{loc}, \nabla\strain, \nabla c)$ is the full set of state
variables Moreover, $\stress_{loc}$, $\nabla\cdot\Stress$ are supposed to be
symmetric second order tensors. The dependence upon $\dot\Gamma_{loc}$ accounts
for possible dissipative contributions to the local stress; we will not consider
dissipative contributions to the non local part of the stress. The balance of
mechanical power is obtained by multiplying the balance of linear momentum
equation by $\vel = \dot{\bm u}$, which leads to
\begin{multline} \label{eqn:balance}
  \frac{d}{dt} \left(\frac{1}{2} \rho \vel^2 \right) =
    [ \nabla\cdot (\stress_{loc} - \nabla\cdot \Stress) + \f] \cdot \vel = \\
    = -(\stress_{loc} : \dot\strain + \Stress \vdots \nabla \dot\strain) +
    \nabla \cdot [\stress \cdot \vel + \Stress : \dot\strain] + \f \cdot \vel,
\end{multline}
having enforced the symmetry condition for the tensors and therefore substituted
$\dot\strain$ with $\nabla\vel$. We interpret (\ref{eqn:balance}) as a balance
equation for the rate of change of kinetic energy and the
\emph{internal and external} (mechanical) powers:
\begin{eqnarray}
  E(\mathcal S) = \int_{\mathcal S} \frac{1}{2} \rho \vel^2 dw,\\
  \power_m^i({\mathcal S}) = \int_{\mathcal S} (\stress_{loc} : \dot\strain +
    \Stress \vdots \nabla \dot\strain)\, dw,\\
  \power_m^e({\mathcal S}) = \int_{\mathcal S} \f \cdot \vel dw +
    \int_{\partial{\mathcal S}} (\stress \cdot \vel +
    \Stress : \dot\strain) \cdot \bm\nu ds,
\end{eqnarray}
where ${\mathcal S}$ is any sub-body, $dw$ is the volume measure, $ds$ the
surface measure and $\bm \nu$ the surface normal. We have used capital letters
for the integrated powers, so that
$\power_m^i({\mathcal S}) = \int_{\mathcal S} \powerden_m^i dw$.

The balance of mechanical power equation is expressed as
\begin{equation}
  \dot E({\mathcal S}) + \power_m^i({\mathcal S}) = \power_m^e({\mathcal S}).
\end{equation}
Now we consider the power associated to the balance of solute mass equation. We
call $\mu$ the \emph{chemical potential}, which is the variable conjugated to
$\dot c$ defining the power balance associated to (\ref{eqn:bal-carbon}):
\begin{equation}\label{eqn:mudotc}
  \mu \dot c = - \mu\nabla \cdot \curr.
\end{equation}
A constitutive expression function of $\Gamma$ for the chemical potential has to
be given, as well as a non-local relation:
\begin{equation}
  \mu = \mu_{loc}(\Gamma_{loc}) - \nabla \cdot \bm M(\Gamma).
\end{equation}
We can rewrite (\ref{eqn:mudotc}) as
\begin{multline} \label{diego}
  \mu(\dot c + \nabla \cdot \curr) = \dot c \mu + \nabla \cdot (\mu\curr) -
    \curr \cdot \nabla \mu = \\
    (\dot c \mu_{loc} + \bm M \cdot \nabla \dot c -
    \curr \cdot \nabla \mu) + \nabla \cdot (\mu \curr - \dot c \bm M) = 0.
\end{multline}
We define the internal and the external powers as follows:
\begin{eqnarray*}
  \power^i_c({\mathcal S}) = \int_{\mathcal S} (\dot c \mu_{loc} +
    \bm M \cdot \nabla \dot c - \curr \cdot \nabla \mu) dw,\\
    \power^e_c({\mathcal S}) = \int_{\partial{\mathcal S}} (\dot c \bm M -
    \mu \curr) \cdot \bm\nu ds
\end{eqnarray*}
so that the balance of powers in the sub-volume ${\mathcal S}$ is expressed as
\begin{equation}
  \power^i_c({\mathcal S}) = \power^e_c({\mathcal S}).
\end{equation}
Note that there is no energy related to a second order time derivative in the
balance of solute mass equation.
%%%%%%%%%%%%%%%%%%%%%%%%%%%%%%%%%%%%%%%%%%%%%%%%%%%%%%%%%%%%%%%%%%%%%%%%%%%%%%%%

\section{Restrictions from the Second Law of Thermodynamics}
\label{second-law}
We will examine the restrictions entailed by the second law of thermodynamics in
the form of the classical Clausius-Duhem inequality:
\begin{equation}
  \dot\ent \geq - \nabla \cdot \frac{\q}{T} + \frac{r}{T}.
\end{equation}
By using the first law of thermodynamics (balance of energy equation):
\begin{equation}\label{eqn:FL}
  \dot\ie = \powerden^i_{tot} - \nabla \cdot \q + r
\end{equation}
(where $\powerden^i_{tot} \equiv \powerden^i_m + \powerden^i_c$ ) to eliminate
the source term $r$, and introducing the \emph{Helmholtz free energy density}
$\fe=\ie-T\ent$, the Clausius-Duhem inequality can be expressed in the more
convenient form (\emph{reduced inequality})
\begin{equation}
  \dot\fe + \dot T \ent - \powerden^i_{tot} \leq - \frac{\q}{T} \nabla T.
\end{equation}
By substituting the expressions for the internal powers and remembering that
$\fe$ is a function of the variables
$\Gamma = (\strain, c, T, \nabla\strain, \nabla c)$, the inequality becomes
\begin{eqnarray*}
  (\fe_T + \ent) \dot T + (\fe_c - \mu_{loc}) \dot c + (\psi_{\nabla c} -
    \bm M) \cdot \nabla \dot c +& & \\
    + (\fe_{\nabla \strain} -
    \Stress) \vdots \nabla \dot\strain + (\fe_{\strain} - \stress_{loc}) : \dot\strain +\curr \cdot \nabla\mu +
      \frac{\q}{T} \cdot \nabla T\leq 0,
\end{eqnarray*}
where the subscripts represent partial derivatives. This inequality has to be
satisfied for all processes
$(\dot\strain, \dot c, \dot T, \nabla \dot\strain, \nabla \dot c,\nabla\mu,\nabla T)$.
Since $\ent$ (as well as $\fe$) depends only on the state, we obtain
\begin{equation}
 \ent = -\fe_T,
\end{equation}
by considering processes with only $\dot T$ different from zero. Similarly, as
also $\mu_{loc}$, $\bm M$, $\Stress$ are assumed state-dependent only, we have
\begin{eqnarray}
  & & \mu_{loc} = \fe_c,\\
  & & \bm M = \fe_{\nabla c},\\
  & & \Stress = \fe_{\nabla\strain}.
\end{eqnarray}
The entropy inequality is therefore reduced to
\begin{equation}
    \bracs{\fe_{\strain} - \stress_{loc}} : \dot\strain
  + \curr \cdot \nabla\mu
  + \frac{\q}{T} \cdot \nabla T \leq 0.
\end{equation}
The current $\curr$ and the heat flux $\q$ are in general functions of the state
and the process. For our purposes it will be sufficient to make the following
simple assumptions ensuring the exploitation of the inequality:
\begin{eqnarray}
  & & \stress_{loc} = \fe_{\strain}+\stress_{an},\qquad \stress_{an}=\bm\gamma:\dot\strain,\\
  & & \curr = -\alpha \nabla \mu,\qquad \alpha(\Gamma) > 0,\\
  & & \q = - \kappa \nabla T,\qquad \kappa(\Gamma) > 0,
\end{eqnarray}
where $\bm\gamma(\Gamma)$ is a positive definite fourth order tensor
of which we provide an explicit formulation in Sect. \ref{par:stress-cp},
$\alpha$ and $\kappa$ are positive coefficients in general dependent on
$\Gamma$, more often only on the temperature $T$. The subscript in the symbol
$\stress_{an}$ stands for \emph{anelastic}. The presence of a dissipative term
in the local stress provides a damping mechanism for the displacive transition,
as showed in Sect.~\ref{res:displ}.

We point out that the chemical potential and the non-dissipative part of the
stress can be written accounting for their local and non-local parts in terms of
functional derivatives of a \emph{free energy functional}
\begin{equation}
 \Fe = \int_{\Omega}\fe(\Gamma)\,dw,
\end{equation}
where $\Omega$ is the material domain and $dw$ the material volume measure. Then,
\begin{eqnarray}
  \mu = \psi_c - \nabla \cdot \psi_{\nabla c} \equiv \frac{\delta\Fe}{\delta c}
    ,\qquad
  \bm\sigma -\stress_{an}= \psi_{\strain}- \nabla \cdot \psi_{\nabla\strain}
    \equiv \frac{\delta\Fe}{\delta\strain},
\end{eqnarray}
these definitions being consistent with the choice of the boundary conditions
made in Sect.\ref{BCs}.

It is possible to substitute the constitutive relations in the equation for the
internal powers, obtaining the following internal power density:
\begin{eqnarray}
  \powerden^i_{tot} = (\fe_{\strain}+ \stress_{an}) : \dot\strain
                    + \fe_{\nabla\strain} \vdots \nabla \dot\strain
                    + \fe_c \dot c
                    + \fe_{\nabla c} \cdot \nabla \dot c
                    + \alpha |\nabla\mu|^2=\\
                    = \dot\fe - \dot T \fe_T + \alpha |\nabla\mu|^2
                    + \dot\strain :(\bm\gamma:\dot\strain).
\end{eqnarray}
Besides, by using $\ent = - \fe_T$, we have
\begin{equation}
  \dot\ie = \frac{d}{dt}(\fe + T \ent) = \dot\fe - \dot T \fe_T - T \dot{\fe_T},
\end{equation}
so that (\ref{eqn:FL}) can be rewritten in term of the free energy as
\begin{equation}\label{eqn:heat}
  - T \dot{\psi_T} = \alpha |\nabla\mu|^2
                   + \dot\strain :(\bm\gamma:\dot\strain)
                   + \nabla \cdot (\kappa \nabla T) + r.
\end{equation}
This is a thermodynamically consistent framework which will be substantiated with
a convenient choice of the free energy $\fe$ (Sect.\ref{free energy}).
%%%%%%%%%%%%%%%%%%%%%%%%%%%%%%%%%%%%%%%%%%%%%%%%%%%%%%%%%%%%%%%%%%%%%%%%%%%%%%%%

\section{Free energy} \label {free energy}
Since constitutive equations have been determined by means of derivatives of the
free energy, the choice of this latter is a convenient way to characterise the
constitutive content of the model. There are two main phenomena involved in the
discussion: the diffusive and the displacive transitions, which have to be
accounted for at the same time; this is done by exploiting the additivity
property of the energy.

The free energy density is written as the sum of different contributions:
\begin{multline} \label {eq2.2}
  \fe = \fe_0(T) + \fe_{thel}(c, e_2, T) + \fe_{diff}(c,T) + \fe_{displ}(\bm\strain,T) \\
      + \fe_{cpl}(c, e_2) + \fe_{grad}(\nabla c, \nabla e_2),
\end{multline}
where:
\begin{itemize}
  \item $\fe_0$ is a term accounting for the background thermal properties of the
    material (contributing to specific heat);
  \item $\fe_{thel}$ is a term accounting for the dilatation due to phase
    transitions and for the contribution of the shear strain;
  \item $\fe_{displ}$, which is a function of $e_2$, accounts for the displacive
    transition;
  \item $\fe_{diff}$, which is a function of $c$, accounts for the diffusive
    transition;
  \item $\fe_{cpl}$ is a coupling term for the interaction between the two phases;
  \item $\fe_{grad}$ is a gradient term and accounts for non-local properties
    and spatial variations of the order parameter.
\end{itemize}
In the Ginzburg-Landau theory of phase transitions, it is assumed that the free energy
density admits a power series expansion with respect to an order parameter.

%-------------------------------------------------------------------------------
\subsection{Diffusive part of the free energy}
The diffusive part of the free energy may be postulated as follows \cite{Cahn00}:
\begin{equation} \label {eq2.3}
  \fe_{diff} = A \dfrac{c^4}{4} + B(T) \dfrac{c^2}{2},
\end{equation}
where $A$ is a constant. The function $B(T)$ is given by:
\begin{equation}
  B(T)=B_0 \dfrac{T-T_P}{T_P},
\end{equation}
with $B_0$ constant and $T_P$ the temperature of perlitic phase transition.
 \begin{figure}
   \center\includegraphics[scale=0.9, trim = 0.3cm 0.5cm 0cm 0cm]{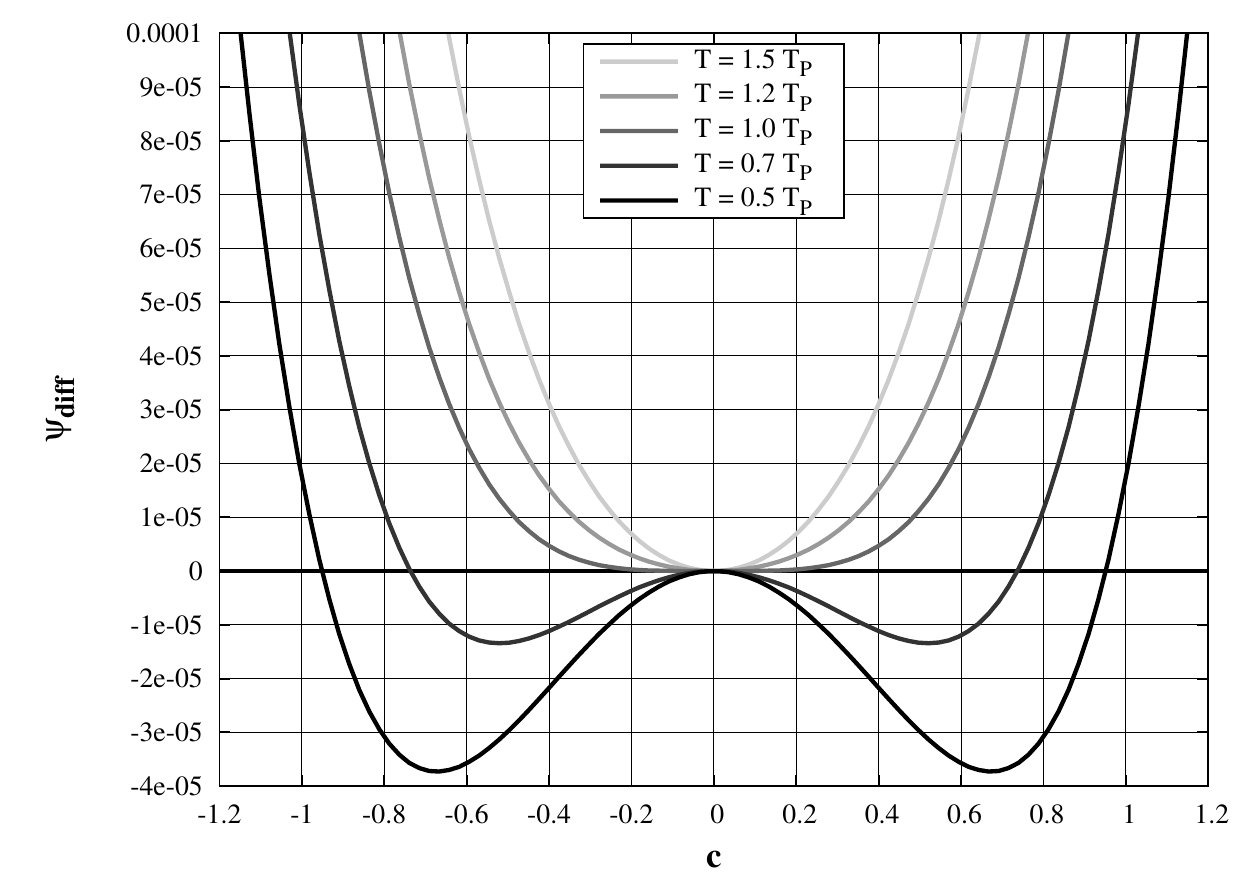}
   \caption{Diffusive free energy.}
 \label{fig:diff_free_energy}
 \end{figure}
In Fig.\ref{fig:diff_free_energy}, $\fe_{diff}$ as a function of $c$ is depicted
for various values of temperature. At high temperature there is only one stable
phase identified as the minimum for $c=0$, while at low temperature a different
stable phase appears for values of $c$ starting from $0$ for $T = T_P$ and
increasing for decreasing temperature. The transition occurs without hysteresis,
i.e. with a continuous variation of the order parameter.

%-------------------------------------------------------------------------------
\subsection{Displacive part of the free energy}
The displacive part of the free energy may be postulated as follows \cite{falk},
\cite{onuki}:
\begin{equation} \label {eq2.6}
  \fe_{displ} = D \dfrac{e_2^6}{6} - E \dfrac{e_2^4}{4}
              + F(T) \dfrac{e_2^2}{2},
\end{equation}
being $D$, $E$ constants. The function $F(T)$ is given by:
\begin{equation}
  F(T) = F_0 \dfrac{T-T_M}{T_M},
\end{equation}
with $F_0$ constant and $T_M$ the temperature of martensitic phase transition.

 \begin{figure}
   \center\includegraphics[scale=0.9, trim = 0.5cm 0.7cm 0cm 0cm]{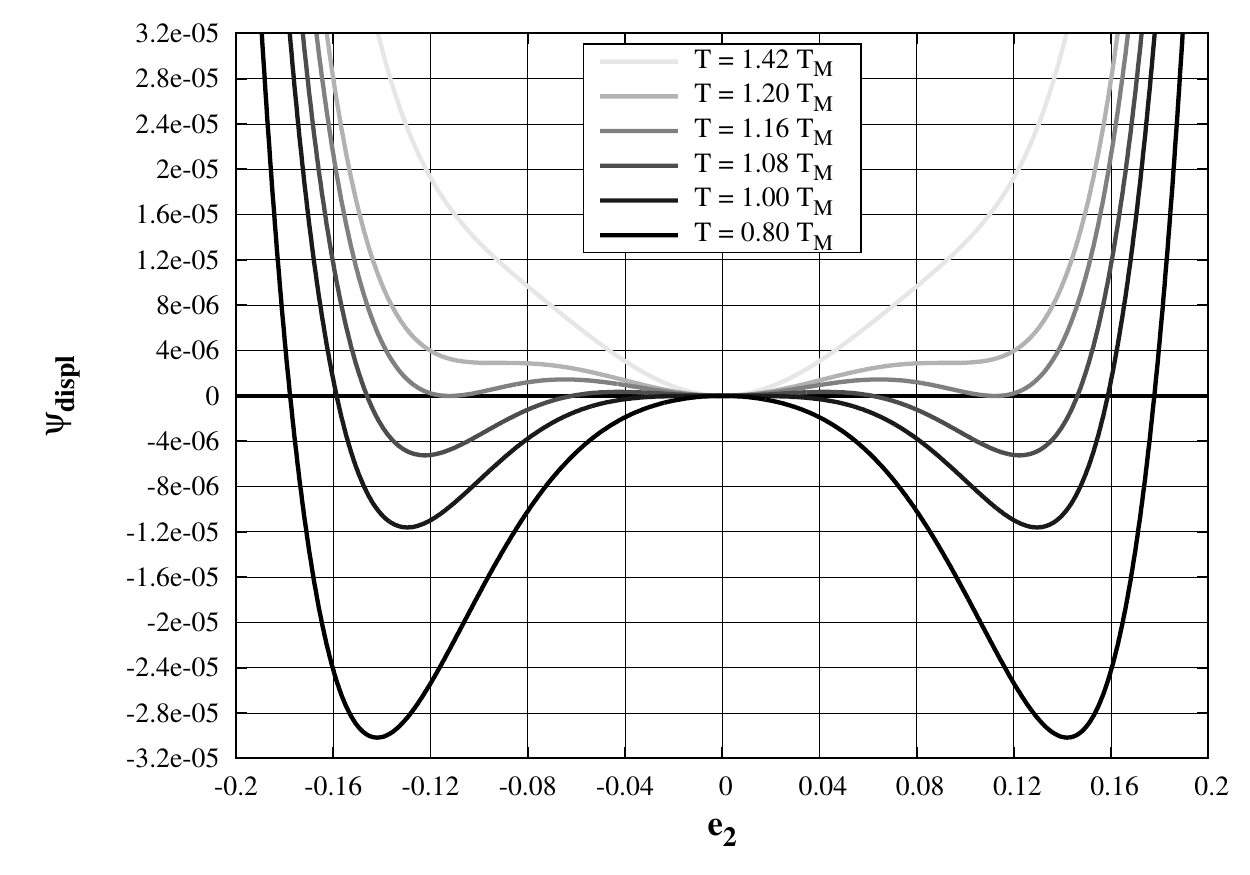}
   \caption{Displacive free energy.}
 \label{fig:displ_free_energy}
 \end{figure}
Fig.\ref{fig:displ_free_energy} shows that at high temperature only one phase
is stable, identified with the minimum for $e_2 = 0$. For $T = T_M$ another
phase become stable, as the free energy shows two minima. Below the critical
temperature, the phase corresponding to the order parameter equal to zero is
unstable, whilst the other one, corresponding to non-zero values of $e_2$, is
stable. Note that, as $e_2$ describes a first order phase change, the transition
between the phases occurs with a jump in the order parameter.

%-------------------------------------------------------------------------------
\subsection{Thermo-elastic part of the free energy}
The expression of $\fe_{thel}$ is the one for the free energy accounting for
the dilatation due to phase transitions \cite{falk} \cite{onuki}, changes in
temperature and for the contribution of the shear strain.
\begin{equation} \label {eq2.7}
  \fe_{thel} = \dfrac{G}{2} \left( e_1 - x_{1c}c - x_{12}e_2^2 - \beta(T-T_0) \right)^2
           + \dfrac{H}{2} e_3^2,
\end{equation}
being $e_1 = \dfrac{1}{\sqrt{2}} \left( \varepsilon_{11} + \varepsilon_{22} \right)$,
$e_3 = \varepsilon_{12}$, $G$, $H$, $T_0$, $x_{1c}$ and $x_{12}$ constants.

%-------------------------------------------------------------------------------
\subsection{Coupling part of the free energy}
\label{nrg:coupl}
$\fe_{cpl}$ is the coupling free energy and penalises the formation of one low
temperature stable phase ($e_2 \neq 0$ or $c \neq 0$) with respect to the other
\cite{falk} \cite{onuki}.
\begin{equation} \label {eq2.8}
  \fe_{cpl} = x_{2c} c^2 e_2^2,
\end{equation}
being $x_{2c}$ a constant.

%-------------------------------------------------------------------------------
\subsection{Gradient part of the free energy}
The gradient part of the free energy is chosen to be the simplest possible one
\cite{falk} \cite{onuki}:
\begin{equation} \label {eq2.9}
  \fe_{grad} = K_e \dfrac{|\nabla e_2|^2}{2} + K_c \dfrac{|\nabla c|^2}{2},
\end{equation}
with $K_e$ and $K_c$ constants; it accounts for non-local effects and spatial
variations of the order parameters (interfaces).
%%%%%%%%%%%%%%%%%%%%%%%%%%%%%%%%%%%%%%%%%%%%%%%%%%%%%%%%%%%%%%%%%%%%%%%%%%%%%%%%

\section{Summary of the equations of the model}
\label{explicit}

\subsection{Constitutive relations: stress and chemical potential}
\label{par:stress-cp}
The constitutive laws of stress and chemical potential can be put in an explicit
form. It is useful to introduce the following notation:
\begin{align}
 &\bm\epsilon_1 = \frac{1}{\sqrt{2}}(\bm i\otimes \bm i +
                                    \bm j\otimes \bm j),\\
 &\bm\epsilon_2 = \frac{1}{\sqrt{2}}(\bm i\otimes \bm i -
                                    \bm j\otimes \bm j),\\
 &\bm\epsilon_3 = \bm i\otimes \bm j + \bm j\otimes \bm i.
\end{align}
The stress tensor can now be written as
\begin{equation}
  \stress = \stress_{loc} - (K_e \nabla^2 e_2)\bm\epsilon_2,
\end{equation}
where the local part is given by four contributions:
\begin{equation}
  \stress_{loc} = \stress_{displ} + \stress_{thel} + \stress_{cpl} + \stress_{an},
\end{equation}
with
\begin{align}
  & \stress_{displ} = \left(D {e_2}^5 - E {e_2}^3
                     + F_0 \frac{T - T_M} {T_M} e_2\right) \,\bm\epsilon_2,\\
  & \stress_{thel}  = G \left( e_1 - x_{1c} c - x_{12} e_2^2 - \beta (T - T_0) \right)
    (\bm\epsilon_1 - 2 x_{12} e_2 \bm\epsilon_2) + H e_3 \bm\epsilon_3,\\
  & \stress_{cpl}   =  2 x_{2c} c^2 e_2 \bm\epsilon_2,
\end{align}
as it regards the conservative contributions; for the dissipative term, we assume
\begin{equation}
\label{eqn:stress-an}
 \stress_{an}=\gamma \dot e_2 \bm\epsilon_2,
\end{equation}
which amounts to consider $\bm\gamma=\gamma \bm\epsilon_2\otimes\bm\epsilon_2$.
In this way, we consider a dissipative term associated only to the displacive
order parameter $e_2$; its effect is to dim the time oscillations in the order
parameter after the displacive transition.

The chemical potential is the following:
\begin{equation}
  \mu = \mu_{loc} - K_c \nabla^2 c,
\end{equation}
where
\begin{multline}
  \mu_{loc} = A c^3 + B_0 \frac{T - T_P}{T_P} c \\
            - G x_{1c} (e_1 - x_{1c} c - x_{12} {e_2}^2 - \beta (T-T_0) )
            + 2 x_{2c} {e_2}^2 c.
\end{multline}
These expressions have to be substituted in the first two equations of system
(\ref{eqn:3bal}).
%-------------------------------------------------------------------------------

\subsection{Equations of the model}
We now summarise the equations of the model. We collect the three balance equations:
\begin{eqnarray}\label{eqn:3bal}
  \left\{\begin{array}{ll}
    \dot c = \nabla\cdot(\alpha \nabla \mu)\\
    \rho \ddot{\bm u} = \nabla \cdot \bm\sigma + \bm b\\
    c_s(T)\dot T - T \left[ \left( \frac{B_0}{T_P} c + G \beta x_{1c} \right) \dot c
        + \left( \frac{F_0}{T_M} + 2 G \beta x_{12} \right) e_2 \dot e_2
        - G \beta \dot e_1 \right] \\
    \qquad \qquad - \alpha |\nabla\mu|^2 -\gamma\dot e_2^2 = \nabla\cdot(\kappa \nabla T) + r,
    \end{array}\right.
\end{eqnarray}
where the last one is the complete heat equation, obtained working out the
expression of $\dot{\fe_T}$ which appears in eq.~(\ref{eqn:heat}) and
accounting for the anelastic stress $\stress_{an}$ in (\ref{eqn:stress-an}).
We remark that
\begin{equation}
c_s(T) = - T\fe_{TT} = - T(\,(\fe_0)_{TT} + G \beta^2 \,)
\end{equation}
denotes the specific heat. The given expression of the heat equation allows to
identify the thermo-chemical and the thermo-mechanical coupling terms
(contained in the second term of eq.~\ref{eqn:3bal}$_3$), which contribute to
the changes in temperature during a phase transition, and the chemical and
mechanical dissipations, respectively $\alpha |\nabla\mu|^2$ and
$\gamma\dot e_2^2$, which act as heat sources when a displacive or a diffusive
transition occurs.

Note that the first of (\ref{eqn:3bal}) is the classical Cahn-Hilliard equation
\cite{cahn}.
%-------------------------------------------------------------------------------

\subsection{Boundary and initial conditions for the differential system}
\label{BCs}
Well-posed boundary and initial conditions are necessary to complement
Eqs.(\ref{eqn:3bal}). Let $\Omega$ be the domain in the material
description (for the purposes exposed here, no distinction is made between the
material particles of the material description and the reference configuration
of the referential description). To state the boundary conditions in a general
mixed form, we assume that the boundary $\partial\Omega$ is partitioned into two
subsets in two different ways:
\begin{equation}
\label{partitions}
  \partial\Omega = \partial\Omega_s \cup \partial\Omega_u
                 = \partial\Omega_q \cup \partial\Omega_T,
\end{equation}
with
\begin{equation}
  \partial\Omega_s \cap \partial\Omega_u = \varnothing , \qquad
  \partial\Omega_q \cap \partial\Omega_T = \varnothing.
\end{equation}
We assume the following boundary conditions:
\begin{eqnarray}\label{eqn:BCs}
  \left\{\begin{array}{ll}
  {\nabla c(\bm x, t) \cdot \bm \nu} \Big |_{\partial\Omega}=0\\
  {\nabla \mu(\bm x, t) \cdot \bm \nu} \Big |_{\partial\Omega}=0\\
  {\nabla e_2(\bm x, t) \cdot \bm \nu} \Big |_{\partial\Omega}=0\\
  {\vect{q} \cdot \bm \nu} \Big |_{\partial\Omega_q} = g(\bm x)\\
  {\stress_{loc}(\bm x, t) \cdot \bm \nu} \Big |_{\partial\Omega_s} = \bm t(\bm x)\\
  {\bm{u}(\bm x, t)} \Big |_{\partial\Omega_u} = \tilde{\bm u}(\bm x)\\
  {T(\bm x, t)} \Big |_{\partial\Omega_T} = \tilde{T}(\bm x).\\
  \end{array}\right.
\end{eqnarray}
We remark that the first and the second conditions ensures the global conservation
of solute mass in the domain:
\begin{equation}
  \frac{d}{dt}\int_{\Omega}c\, dw=\int_{\partial\Omega}\curr\cdot\bm\nu\,ds=0.
\end{equation}
The non-local boundary conditions (Eqs.(\ref{eqn:BCs})$_2$, (\ref{eqn:BCs})$_3$)
are needed in this model due to the non-local form of the stress
\cite{Polizzotto}. The other conditions are classical.

The initial conditions are the following:
\begin{eqnarray}\label{eqn:ic}
  \left\{\begin{array}{ll}
    \bm{u}(\bm x, 0)=\bm u_0 (\bm x)\\
    \dot{\bm{u}}(\bm x, 0)=\bm v_0 (\bm x)\\
    c(\bm x, 0)=c_0(\bm x)\\
    T(\bm x, 0)=T_0(\bm x).
  \end{array}\right.
\end{eqnarray}

\section{Conclusions}
A model to study both diffusive and displacive phase transitions has been
proposed in a thermodynamically consistent framework in which, for the presence
of the balance of energy equation, non-isothermal conditions can be accounted
for. The resulting model is a diffuse interface one, and relies on the choice of
two order parameters and of its gradients; this rises some issues about the
exploitation of the Clausius-Duhem inequality, and a way to deal with this has
been proposed. In particular, we have considered a suitable modification of the
first law of thermodynamics by including non-conventional expressions for the
internal powers (due to non-localities), keeping at the same time the classical
Clausius-Duhem form for the second law of thermodynamics.

The equations of the model are the balance of linear momentum equation, the
Cahn-Hilliard equation and the heat equation; the first two are the evolution
equations for the two order parameters; the heat equation accounts for thermal
effects like cooling or heat released due to a phase change. The model is
complemented by a proper description of the free energy and consistent boundary
and initial conditions; due to the presence of the gradients of the order
parameters, two higher order boundary conditions appear to be necessary, and
this can be also observed in the exploitation of the Clausius-Duhem inequality.

% Some numerical examples have been presented that show how the model is able
% to describe the diffusive and the displacive phase transitions as well as the
% accompanying thermo-chemo-mechanical effects, likewise the change in
% temperature induced by the transitions and the interplay between the different
% phases.

%%%%%%%%%%%%%%%%%%%%%%%%%%%%%%%%%%%%%%%%%%%%%%%%%%%%%%%%%%%%%%%%%%%%%%%%%%%%%%%%
\section*{Acknowledgements}
The authors wish to thank Prof. Mauro Fabrizio, Prof. Pier Gabriele Molari and
Prof. Francesco Ubertini for the very useful discussions and the University of
Bologna for the financial support.

%%%%%%%%%%%%%%%%%%%%%%%%%%%%%%%%%%%%%%%%%%%%%%%%%%%%%%%%%%%%%%%%%%%%%%%%%%%%%%%%

\end{document}